# A proposal for a feasible quantum-optical experiment to test the validity of the No-Signaling Theorem

Demetrios A. Kalamidas

*Raith USA*, *2805 Veterans Memorial Hwy, Ronkonkoma, NY, 11779*

*Email:* dakalamidas@sci.ccny.cuny.edu   *Telephone: 1-631-738-9500   Fax:  1-631-738-2055*

**Abstract**

   Motivated by a proposal from Greenberger [*Physica Scripta* **T76**, p.57 (1998) ] for superluminal signaling, and inspired by an experiment from Zou, Wang, and Mandel [*Phys. Rev. Lett.* **67**, p.318 (1991) ] showing interference effects within multi-particle entanglement without coincidence detection, we propose a feasible quantum-optical experiment that purports to manifest the capacity for superluminal transfer of information between distant parties.

**Introduction**

 Numerous experiments to date, mainly in the quantum-optical domain, seem to strongly support the notion of an inherent nonlocality pertaining to certain multi-particle quantum mechanical processes. However, with apparently equal support, this time from a theoretical perspective, it is held that these nonlocal 'influences' cannot be exploited to produce superluminal transfer of information between distant parties. The theoretical objection to superluminal communication, via quantum mechanical multi-particle entanglement, is essentially encapsulated by the 'no-signaling theorem' [1].  So, it is within this context that we present a scheme whose mathematical description leads to a result which directly contradicts the no-signaling theorem and manifests, using only the standard quantum mechanical formalism, the capacity for superluminal transmission of information.

The pursuit of such a scheme was motivated by a superluminal signaling set-up proposed by Greenberger [2], because it seemed to employ only standard quantum mechanical transformations and called into question the generality of the no-signaling theorem. To our knowledge, Greenberger's proposal has only very recently been challenged, for the first time, by Ghirardi and Romano [3]. Greenberger's scheme requires a device that is, according to him, possible in principle but not currently available in the technological sense. The set-up proposed here can be realized with existing technology and application of well-known techniques in quantum optics, although serious technical difficulties will have to be overcome, due to the fragile nature of maintaining quantum coherence over large distances.

The authors of [3] also challenged the validity of the scheme that will be described here, declaring that the same arguments used to attack Greenberger's claims can be applied to the claims of this paper. However, the authors of [3] have subsequently retracted their declaration [4] and have conceded that the proposal described here is fundamentally different from Greenberger's and cannot be challenged by the arguments they presented in [3].

Lastly, the actual idea for this scheme was inspired by the truly striking experiment of Zou, Wang, and Mandel [5], wherein single-photon interference was extracted from entangled photon pairs *without* coincidence detection. In [5] one could send signals in a purely quantum mechanical manner, by creating or destroying the conditions for single-photon interference, but space-like separation between the communicating parties is impossible for that set-up. We will show that the scheme described here does allow space-like separation between the communicating parties, thus enabling superluminal signaling, yet it is also based on creating or destroying the conditions for single-photon interference.

# Experimental Proposal

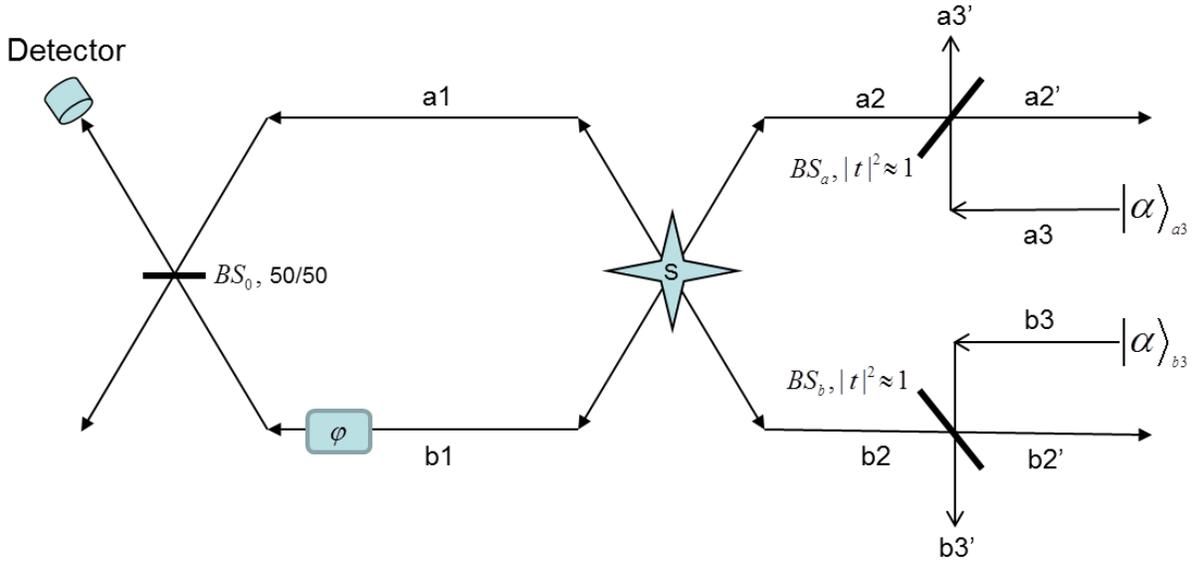

Consider the illustration above. A source S creates single-pair, maximally path-entangled photons: a photon pair can be emitted into modes **a1a2** or into modes **b1b2**, with equal probability. This entangled state can be expressed as:

$$|\psi\rangle_S = \frac{1}{\sqrt{2}}\left(|1\rangle_{a1}|1\rangle_{a2}|0\rangle_{b1}|0\rangle_{b2} + e^{i\varphi}|0\rangle_{a1}|0\rangle_{a2}|1\rangle_{b1}|1\rangle_{b2}\right) \quad (1).$$

In modes **a3** and **b3** there are two input coherent states, $|\alpha\rangle_{a3}$ and $|\alpha\rangle_{b3}$ respectively, having equal amplitudes and phases. Modes **a1** and **b1** combine at a 50/50 beam splitter, **BS$_0$**. Modes **a2** and **b2** combine with modes **a3** and **b3**, respectively, at beam splitters **BS$_a$** and **BS$_b$**. It is stipulated that these two beam splitters are of a 'high-transmission' type so that: if 't' and 'r' denote the complex transmission and reflection coefficients respectively, with $|t|^2+|r|^2=1$, we will have t→1 and r→0.

We assume that the two coherent states are spectrally matched to and have the same polarization as the photons from the entangled pairs. Furthermore, we assume that the transmitted and reflected components of these states, at the high-transmission beam splitters, have good spatial and temporal overlap with those of the photons from the entangled pairs.

At **BS$_a$** and **BS$_b$**, we note the following transformations:

$$|1\rangle_{a2}|\alpha\rangle_{a3} \xrightarrow{BS_a} \left(t\hat{a}^\dagger_{a2'} + r\hat{a}^\dagger_{a3'}\right)|t\alpha\rangle_{a3'}|r\alpha\rangle_{a2'} \quad (2a)$$

$$|1\rangle_{b2}|\alpha\rangle_{b3} \xrightarrow{BS_b} \left(t\hat{a}^\dagger_{b2'} + r\hat{a}^\dagger_{b3'}\right)|t\alpha\rangle_{b3'}|r\alpha\rangle_{b2'} \quad (2b),$$

which we will use in calculating the total output state.

The initial amplitudes of the coherent states are adjusted so that their tiny reflected components (at **BS$_a$** and **BS$_b$**) satisfy the condition of very weak coherent states, containing at most one photon:

$$|r\alpha\rangle_{a2'} \approx |0\rangle_{a2'} + r\alpha|1\rangle_{a2'} = \left(1 + r\alpha\hat{a}^\dagger_{a2'}\right)|0\rangle_{a2'} \quad (3a)$$

$$|r\alpha\rangle_{b2'} \approx |0\rangle_{b2'} + r\alpha|1\rangle_{b2'} = \left(1 + r\alpha\hat{a}^\dagger_{b2'}\right)|0\rangle_{b2'} \quad (3b).$$

If we suppose that the source S creates the entangled state $|\psi\rangle_S$ with an amplitude $\varepsilon$, then the initial total state, $|\Psi\rangle_0$, may be written as follows:

$$|\Psi\rangle_0 = \varepsilon|\psi\rangle_S \otimes |\alpha\rangle_{a3} \otimes |\alpha\rangle_{b3} =$$

$$= \frac{\varepsilon}{\sqrt{2}}\left(|1\rangle_{a1}|1\rangle_{a2}|0\rangle_{b1}|0\rangle_{b2} + e^{i\varphi}|0\rangle_{a1}|0\rangle_{a2}|1\rangle_{b1}|1\rangle_{b2}\right) \otimes |\alpha\rangle_{a3} \otimes |\alpha\rangle_{b3} \quad (4).$$

After beam splitters **BS$_a$** and **BS$_b$**, using {(2a),(2b)} and {(3a),(3b)}, we have:

$$|\Psi\rangle_0 \xrightarrow[BS_b]{BS_a} \frac{\varepsilon}{\sqrt{2}} \Big[|1\rangle_{a1}|0\rangle_{b1}\big(t\hat{a}^\dagger_{a2'} + r\hat{a}^\dagger_{a3'}\big) + e^{i\varphi}|0\rangle_{a1}|1\rangle_{b1}\big(t\hat{a}^\dagger_{b2'} + r\hat{a}^\dagger_{b3'}\big)\Big] \cdot$$

$$\cdot \big(1 + r\alpha\hat{a}^\dagger_{a2'}\big)\big(1 + r\alpha\hat{a}^\dagger_{b2'}\big)|0\rangle_{a2'}|0\rangle_{b2'}|t\alpha\rangle_{a3'}|t\alpha\rangle_{b3'} = \quad (5a)$$

$$= \frac{\varepsilon}{\sqrt{2}}\Big[t\big(|1\rangle_{a1}|0\rangle_{b1}\hat{a}^\dagger_{a2'} + e^{i\varphi}|0\rangle_{a1}|1\rangle_{b1}\hat{a}^\dagger_{b2'}\big) + \underbrace{r\big(|1\rangle_{a1}|0\rangle_{b1}\hat{a}^\dagger_{a3'} + e^{i\varphi}|0\rangle_{a1}|1\rangle_{b1}\hat{a}^\dagger_{b3'}\big)}_{negligible}\Big] \cdot$$

$$\cdot \big(1 + r\alpha\hat{a}^\dagger_{a2'} + r\alpha\hat{a}^\dagger_{b2'} + r^2\alpha^2\hat{a}^\dagger_{a2'}\hat{a}^\dagger_{b2'}\big)|0\rangle_{a2'}|0\rangle_{b2'}|t\alpha\rangle_{a3'}|t\alpha\rangle_{b3'} \approx \quad (5b)$$

$$\approx \frac{\varepsilon t}{\sqrt{2}}\big(|1\rangle_{a1}|0\rangle_{b1}\hat{a}^\dagger_{a2'} + e^{i\varphi}|0\rangle_{a1}|1\rangle_{b1}\hat{a}^\dagger_{b2'}\big)\big(1 + r\alpha\hat{a}^\dagger_{a2'} + r\alpha\hat{a}^\dagger_{b2'} + r^2\alpha^2\hat{a}^\dagger_{a2'}\hat{a}^\dagger_{b2'}\big)|0\rangle_{a2'}|0\rangle_{b2'} \cdot$$

$$\cdot |t\alpha\rangle_{a3'}|t\alpha\rangle_{b3'} \equiv |\Psi\rangle_{out} \quad (5c).$$

From (5c) we have discarded the underlined part in (5b), which is proportional to 'r' and whose two terms represent the highly unlikely event of a right-going photon, from an entangled pair, being reflected at one or the other of the high-transmission beam splitters, **BS$_a$** or **BS$_b$**. Such events are rare because r→0, making the contribution of these two terms, to the form of the total output state, negligible. In contrast, terms proportional to 'rα' are significant because 'α' can be arbitrarily large in magnitude. So, after discarding these terms, applying the creation operators to their respective kets, and re-arranging the terms, the final state can be written, to good approximation, as:

$$|\Psi\rangle_{out} = \frac{\varepsilon t}{\sqrt{2}} \Big[ \underbrace{r\alpha \big(|1\rangle_{a1}|0\rangle_{b1} + e^{i\varphi}|0\rangle_{a1}|1\rangle_{b1}\big)|1\rangle_{a2'}|1\rangle_{b2'}}_{\text{interference part}} + |1\rangle_{a1}|0\rangle_{b1}|1\rangle_{a2'}|0\rangle_{b2'} +$$

$$+ e^{i\varphi}|0\rangle_{a1}|1\rangle_{b1}|0\rangle_{a2'}|1\rangle_{b2'} + \sqrt{2}r\alpha\big(|1\rangle_{a1}|0\rangle_{b1}|2\rangle_{a2'}|0\rangle_{b2'} + e^{i\varphi}|0\rangle_{a1}|1\rangle_{b1}|0\rangle_{a2'}|2\rangle_{b2'}\big) +$$

$$+ \sqrt{2}r^2\alpha^2\big(|1\rangle_{a1}|0\rangle_{b1}|2\rangle_{a2'}|1\rangle_{b2'} + e^{i\varphi}|0\rangle_{a1}|1\rangle_{b1}|1\rangle_{a2'}|2\rangle_{b2'}\big) \Big] |t\alpha\rangle_{a3'}|t\alpha\rangle_{b3'} \qquad (6).$$

From (6) we notice that only the underlined part, namely

$$r\alpha\big(|1\rangle_{a1}|0\rangle_{b1} + e^{i\varphi}|0\rangle_{a1}|1\rangle_{b1}\big)|1\rangle_{a2'}|1\rangle_{b2'},$$

includes a coherent superposition of single-photon occupation possibilities between modes **a1** and **b1**. This part alone will lead to single-photon interference at **BS₀** and corresponds to the outcome $|1\rangle_{a2'}|1\rangle_{b2'}$: there is one photon in mode **a2'** and one photon in mode **b2'**. The coherent superposition arises because of the *fundamental indistinguishability* of the following two possibilities corresponding to the $|1\rangle_{a2'}|1\rangle_{b2'}$ outcome:

*Did the photon in mode **a2'** derive from the entangled pair, while the photon in mode **b2'** derived from the weak coherent state $|r\alpha\rangle_{b2'}$?*

*(Implying that a photon exists in mode **a1**)*

OR

*Did the photon in mode **a2'** derive from the weak coherent state $|r\alpha\rangle_{a2'}$, while the photon in mode **b2'** derived from the entangled pair?*

*(Implying that a photon exists in mode **b1**).*

For the other possible combinations of photon occupation in modes **a2'** and **b2'** we can determine, in principle, if the photon propagating towards **BS$_0$** exists in mode **a1** or **b1**, thus destroying the condition for single-photon interference at that beam splitter, since we have 'which-way' information for that photon. For instance, from $|\Psi\rangle_{out}$ above, let us analyze the outcome $|1\rangle_{a1}|0\rangle_{b1}|1\rangle_{a2'}|0\rangle_{b2'}$ : since there is one photon in mode **a2'** and zero photons in mode **b2'**, we can deduce that both coherent states were empty and that the single photon in mode **a2'** derived from the entangled pair, thus revealing that a photon exists in mode **a1**. As a further example, let us analyze the outcome $|0\rangle_{a1}|1\rangle_{b1}|1\rangle_{a2'}|2\rangle_{b2'}$ : since there is one photon in mode **a2'** and two photons in mode **b2'**, we can deduce that both coherent states contributed their maximum of one photon each, and that the extra photon in mode **b2'** derived from the entangled pair, thus revealing that a photon exists in mode **b1**.

In other words, when the coherent states are present, their tiny reflected components will emerge in the same two spatio-temporal modes as those of the transmitted right-going photons from the entangled pairs and, when the $|1\rangle_{a2'}|1\rangle_{b2'}$ outcome occurs, this fact leads to *quantum erasure* of the 'which-way' information for a corresponding left-going photon propagating towards **BS$_0$**. Conversely, in the final expression for $|\Psi\rangle_{out}$, except for the underlined part, the other six terms represent outcomes that project a left-going photon onto a definite spatial mode and therefore, for those outcomes, single-photon interference (at the output ports of **BS$_0$**) will not be present.

**Implication**

The most important implication of the scenario described in this paper can be illustrated as follows: Even if one did not perform any measurements on any of the right-going photons emerging from modes **a2'** and **b2'**, a subset of left-going photons, corresponding to the $|1\rangle_{a2'}|1\rangle_{b2'}$ outcome, will exhibit interference (at the output ports of **BS$_0$**) because this outcome occurs with probability $\frac{1}{2}|\varepsilon tr\alpha|^2$ regardless of measurement. This means that one would obtain *low-visibility* single-photon interference, at the output ports of **BS$_0$**, over the entire set of runs, without coincidence measurements. The low visibility is due to the presence of the other non-interfering terms in the final expression for $|\Psi\rangle_{out}$. On the other hand, if the coherent states were absent, it would be the case that 'which-way' information for a left-going photon would be available for all runs and, therefore, single-photon interference would *not occur at all* at the output ports of **BS$_0$** : without the coherent states the only possible outcomes in modes **a2'** and **b2'** would be $|1\rangle_{a2'}|0\rangle_{b2'}$ and $|0\rangle_{a2'}|1\rangle_{b2'}$, which allow one to infer the mode of a corresponding left-going photon, with certainty, for all runs.

A direct consequence of the above argument is that superluminal signaling becomes possible, this being accomplished by switching the coherent states 'ON' and 'OFF' in an appropriate manner. One communication method could be as follows: The receiver is on the left wing of the set-up and monitors the photon counts on one of the output ports of **BS$_0$**. He adjusts the phase φ to a value resulting in maximal constructive single-photon interference at that output port. The sender is on the right wing of the set-up and controls the switching of the coherent states. Assuming the source S creates entangled photon pairs at some known rate, he can choose to switch the coherent states 'ON' and 'OFF' for fixed time intervals, in a sequential pattern, and thus form a binary code. The receiver will note the counts (on his chosen output port of **BS$_0$**) for each time interval and interpret the readings. If the reading for a particular time interval is greater than 50% of the expected total rate, he infers that the coherent states are 'ON', since the higher count is a result of constructive interference at that output port (due to the effect described thus far). If the reading is 50% of the expected total rate, he infers that the coherent states are 'OFF', since there is no longer any interference effect and each left-going photon can occupy one of the two output ports of **BS$_0$** with equal probability. Thus, the coherent states being 'ON' and 'OFF' can represent the '1' and '0' bit values, respectively.

**Summary and Conclusion**

In summary, we have proposed a quantum-optical experiment that purports to enable the superluminal transfer of information, via the fundamental nonlocality presumed to be inherent to multi-particle quantum entanglement. The outcome of the quantum mechanical description of the scheme that we have presented here provides a clear counter-example to the no-signaling theorem, questioning its generality.

As far as the feasibility of the scheme is concerned, it would indeed be technically challenging, yet all of the elements that make it up have been experimentally demonstrated already: creation of single pairs of path-entangled photons, along with single-photon detection, within interferometrically stable configurations [6], mixing of weak coherent states with single-photon Fock states [7], and long-distance implementation of quantum communication protocols [8].

In conclusion, we note that the quantum mechanical description of the scheme presented here may entail the following: an empirical 'null result' (the absence of superluminal signals) could have serious consequences regarding the viability of the notion of 'nonlocality' as the underlying cause of certain types of correlation within multi-particle entangled quantum states, or point to the inadequacy of the standard quantum mechanical formalism in providing a fully accurate description of certain physical situations ([9], [10]).